\documentclass[twocolumn,prl,aps,10pt]{revtex4-1}

\usepackage{amsfonts,amssymb}
\usepackage[sumlimits,intlimits]{amsmath}
\usepackage{graphics}
\usepackage{graphicx}
\usepackage{mathrsfs}
\usepackage{textcomp}
\usepackage{verbatim}
\usepackage{hyperref}    
\usepackage{bm}          

\newcommand{\be}{\begin{equation}}
\newcommand{\ee}{\end{equation}}
\newcommand{\Tes}{T_\text{ES}}
\newcommand{\e}{\varepsilon}

\begin{document}

\title{Why is the bulk resistivity of topological insulators so small?}

\author{Brian Skinner}
\author{Tianran Chen}
\author{B. I. Shklovskii}
\affiliation{Fine Theoretical Physics Institute, University of Minnesota, Minneapolis, MN 55455, USA}

\date{\today}

\begin{abstract}
As-grown topological insulators (TIs) are typically heavily-doped $n$-type crystals. Compensation by acceptors is used to move the Fermi level to the middle of the band gap, but even then TIs have a frustratingly small bulk resistivity.  We show that this small resistivity is the result of band bending by poorly screened fluctuations in the random Coulomb potential. Using numerical simulations of a completely compensated TI, we find that the bulk resistivity has an activation energy of just $0.15$ times the band gap, in good agreement with experimental data. At lower temperatures activated transport crosses over to variable range hopping with a relatively large localization length.
\end{abstract}

\maketitle

The three-dimensional (3D) topological insulator (TI)~\cite{Fu2007tii, Moore2007tio, Roy2009tpa, Fu2007tiw, Qi2008tft} has gapless surface states that are expected to exhibit a range of interesting quantum phenomena~\cite{Hasan2010c:t, Qi2011tia}.  While a number of 3D TIs have been identified, most of these are poor insulators in the bulk, so that the properties of the surface are obscured in transport measurements.  For this reason achieving a bulk-insulating state remains an active topic of research~\cite{Qu2010qoa, Analytis2010tss, Xiong2012hsh, Checkelsky2009qii, Butch2010sss, Analytis2010bfs, Eto2010aoo, Ren2011oot, Ren2011o$s, Ren2012flt}. 

Typically, as-grown TI crystals are heavily doped $n$-type semiconductors, and correspondingly they exhibit metallic conduction. In order to make them insulating these TIs are compensated by acceptors. With increasing compensation $K = N_A/N_D$, where $N_D$ and $N_A$ are the concentrations of monovalent donors and acceptors, repsectively, the Fermi level shifts from the conduction band to inside the band gap and then at $K > 1$ into the valence band. When compensation of donors is complete, $K=1$, the Fermi level is in the middle of the gap and the most insulating state of the TI is reached.  The hope is that at $K = 1$ the bulk resistivity $\rho$ should obey the activation law, 
\be
\rho = \rho_0 \exp(\Delta/k_BT),
\label{act}
\ee
with an activation energy $\Delta$ that is equal to half the band gap $E_g$.  Here,  $\rho_0$ is a constant and $k_BT$ is the thermal energy.  Since typically $E_g \sim 0.3$ eV, this expectation would imply a well-insulating bulk at room temperatures and below.

The typical experimental situation at $K=1$, however, is frustrating~\cite{Ren2011o$s}. In the range of temperatures between $100$ and $300$ K the resistivity is activated, but with an activation energy that is three times smaller than expected, $\Delta \sim 50$ meV.  At $T \lesssim 100$ K the activated transport is replaced by variable range hopping (VRH), characterized by $\rho \propto \exp[(T_0/T)^x]$ with $x < 1$, and the resistivity grows even more slowly with decreasing $T$.  In Ref.\ \onlinecite{Ren2011o$s} the authors show that Mott VRH ($x = 1/4$) provides a reasonable fit to their data at $50$ K $\lesssim T \lesssim 100$ K.  Definite characterization of the temperature exponent $x$ is difficult, however, due to the relatively narrow window of temperature and to variations between samples.  At $T \lesssim 50$ K the resistivity saturates due to the contribution of the surface states.

\begin{figure}[htb!]
\centering
\includegraphics[width=0.45 \textwidth]{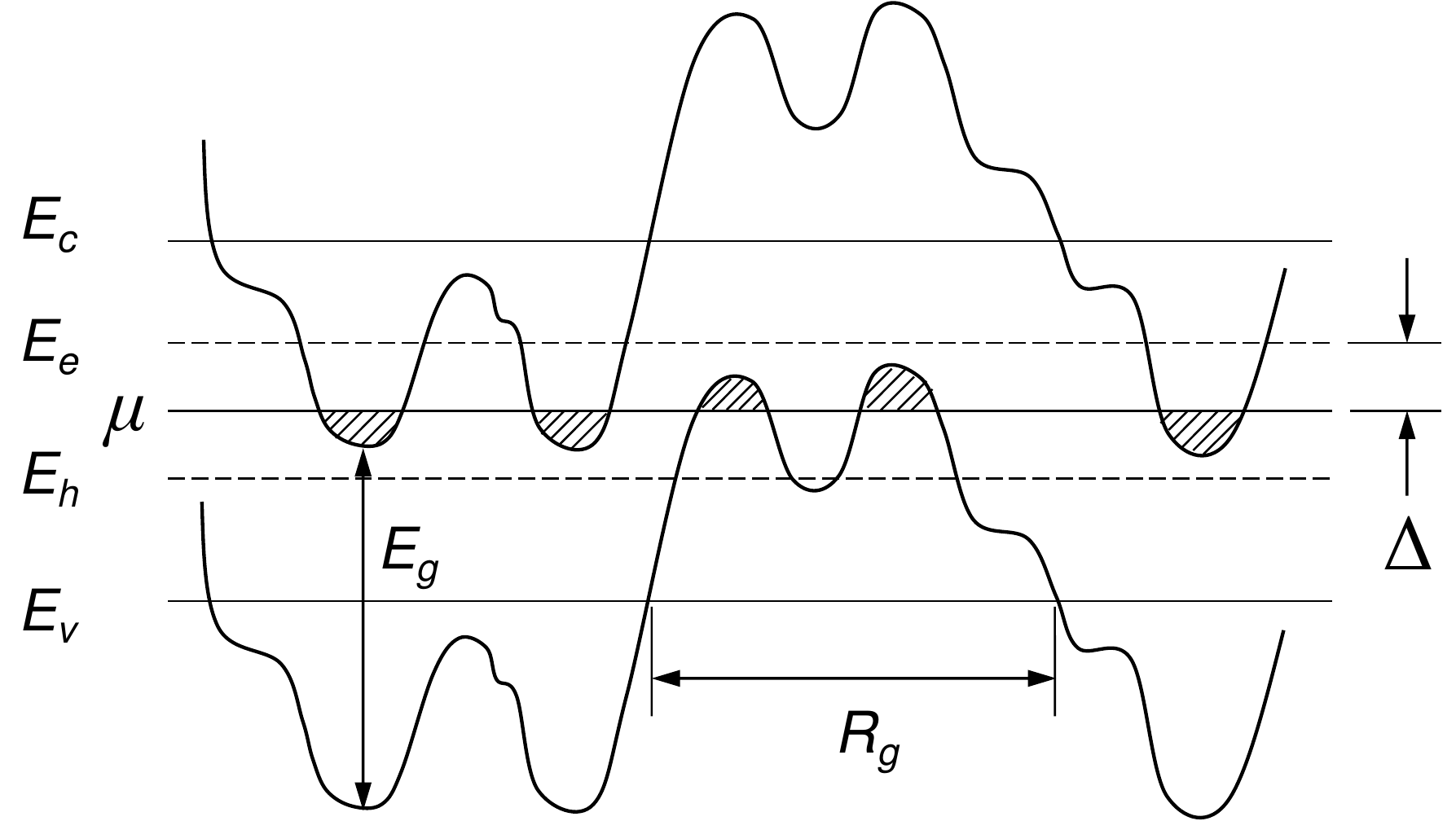}
\caption{Energy diagram of a completely compensated TI with band gap $E_g$. The upper and  the lower straight lines ($E_c$ and $E_v$) indicate the unperturbed positions of the bottom of the conduction band and the ceiling of the valence band; the middle line ($\mu$) corresponds to the Fermi level. Meandering lines represent the band edges, which are modulated by the fluctuating potential of charged impurities; $R_g$ is the characteristic size of these potential fluctuations.  The percolation levels for electrons, $E_e$, and holes, $E_h$, are shown by dashed lines; the activation energy $\Delta$ corresponds to the difference $E_e - \mu$ (or $\mu - E_h$). Puddles occupied by carriers are shaded. Shallow impurity levels are not shown because they merge with the band edges.} \label{fig:bands}
\end{figure}

In this paper we suggest an explanation for the unexpectedly small bulk resistivity of TIs. We assume that both donors and acceptors are shallow and we use the theory of completely compensated semiconductors (CCS)~\cite{Shklovskii1972ccc, Efros1984epo}. This theory is based on the idea that near $K=1$, when almost all donors and acceptors are charged, random fluctuations in the local concentration of impurities result in large fluctuations of charge. The resulting Coulomb potential is poorly screened because of the vanishing average concentration $n = N_D - N_A$ of screening electrons. Huge fluctuations in the random potential bend the conduction and valence bands edges and in some places bring them to the Fermi level, thereby creating electron and hole puddles that non-linearly screen the random potential. Thus, the amplitude of fluctuations is limited only by the semiconductor gap $E_g$. As a result the ground state of a CCS, shown in Fig. \ref{fig:bands}, is similar to a network of $p$-$n$ junctions~\cite{Shklovskii1972ccc, Efros1984epo}.  The characteristic size of these $p$-$n$ junctions, also called the nonlinear screening radius, is given by 
\be
R_g = \frac {E_g^{2}\kappa^2}{8\pi Ne^{4}},
\label{Rg}
\ee 
where $\kappa$ is the dielectric constant, $e$ is the electron charge, and $N = N_D = N_A$. For $N = 10^{19}$ cm$^{-3}$ and $\kappa = 20$, $R_g \approx 70$ nm $\gg N^{-1/3} \approx 4.6$ nm, so that we deal with a very long range potential. As a result, the resistivity can be dramatically different from the expectation outlined above, which assumed flat bands. First, at relatively high temperatures conduction is due to electrons and holes being activated from the Fermi level to their corresponding classical percolation levels (classical mobility edges), $E_e$ and $E_h$, in the conduction and the valence bands.  These may be substantially closer to the Fermi level $\mu$ than $E_g/2$, but so far the resulting value of $\Delta$ has not been studied theoretically. Second, at sufficiently low temperatures electrons and holes can hop (tunnel) between distant puddles, so that variable range hopping replaces activated transport.  In the low temperature limit $\rho(T)$ should obey the Efros-Shklovskii (ES) law of VRH ~\cite{Efros1975cga},
\be
\rho = \rho_{0}\exp\left[ (\Tes/T)^{1/2} \right],
\label{eslaw}
\ee
where $\Tes = Ce^2/\kappa\xi$, $\xi$ is the localization length of states with energy close to the Fermi level, and $C$ is a numerical coefficient.  So far the magnitude of $\Tes$ and the nature of the crossover between activated and VRH conduction have not been studied.

In this paper, motivated by the TI resistivity puzzle, we return to CCS and model numerically the $K = 1$ case. For moderately large $T$ we find that $\Delta = 0.15 E_g$. For a TI with $E_g = 0.3$ eV this implies $\Delta = 45$ meV, in agreement with observed values~\cite{Ren2011o$s}. We also calculate the single-particle density of states (DOS) of impurity states, and we find that the DOS has a Coulomb gap at the Fermi level~\cite{Efros1975cga}. We show from our simulation that the resistivity is described by Eq.\ (\ref{eslaw}) at low temperatures and crosses over to Eq.\ (\ref{act}) at higher $T$. We present a crude estimate of the localization length $\xi$ which suggests that $\Tes \sim 900$ K and that the crossover between activation and ES VRH occurs at $T \sim 40$ K.  Together our results for the activated and VRH resistivity establish a universal upper limit for the resistivity $\rho(T)$ that one can achieve for a 3D TI compensated by shallow inpurities. 

In order to model the CCS numerically, we simulate a cube filled by an equal number of randomly positioned donors and acceptors (20000 of each).  We numerate all donors and acceptors by the index $i$ and we define $n_i = 0, 1$ as the number of electrons residing at impurity $i$ and the variable $f_i$ to discriminate between donors ($f_i = 1$) and acceptors ($f_i = -1$). The resulting Hamiltonian is 
\be
H = \frac{E_g}{2} \sum_i f_i n_i +  \sum_{\langle ij \rangle} V(r_{ij}) q_i q_j,
\label{Hamiltonian}
\ee
where $q_i = (f_i + 1)/2 - n_i$ is the net charge of site $i$ and all energies are defined relative to the middle of the band gap. The first term of Eq.\ (\ref{Hamiltonian}) contains the energies of donor and acceptor sites, which for the case of shallow impurities is very close to $\pm E_g/2$.
The second term is the total interaction energy of charged impurities.  For two impurities at a distance $r \gg a_B$, where $a_B$ is the Bohr radius of impurity states, one can describe their interaction using the normal Coulomb law $V(r) = e^2/\kappa r$. For example, an empty donor shifts the energy of an electron at a distant filled donor by an amount $- e^2/\kappa r$.  On the other hand, for a pair of impurities with separation $r < a_B$, quantum mechanical averaging over the electron wavefunction becomes important (such close impurity pairs are common in heavily doped semiconductors, where $a_B > N^{-1/3}$).  A pair of very close donors, for example, cannot create an electron state deeper than that of the helium-like ion with binding energy $2e^2 /\kappa a_B$. In order to capture this quantum phenomenon in an approximate way, we use the classical Hamiltonian of Eq.\ (\ref{Hamiltonian}) with a truncated Coulomb potential $V(r)=e^2/\kappa (r^2 + a_B^2)^{1/2}$.  The result of this truncation is to eliminate the unphysically deep electron states that would result from very compact impurity pairs with an unmodified $1/r$ interaction.  We will show below that our results are mostly insensitive to the details of this truncation.  Note that Eq.\ (\ref{Hamiltonian}) does not include the kinetic energy of electrons and holes in the conduction and valence bands and, therefore, aims only at describing the low temperature ($k_B T \ll E_g$) physics of CCS. 

In all results below we use dimensionless units for $r$, $a_B$, $\xi$, $H$, $E_g$, and $k_B T$, measuring all distances in units of $N^{-1/3}$ and all energies in units of $e^2 N^{1/3}/\kappa$. Thus, Eq.\ (\ref{Hamiltonian}) can be understood as dimensionless, with  $E_g \gg 1$ and $V(r) = (r^2 + a_B^2)^{-1/2}$. For a TI with $E_g = 0.3$ eV, $\kappa = 20$ and $N=10^{19} $ cm$^{-3}$, the unit of energy $e^2 N^{1/3}/\kappa \approx 15$ meV, so that the dimensionless gap $E_g \approx 20$.  We were unable to directly model $E_g = 20$, since in this case the very large $R_g \approx 16$ leads to large size effects. Instead, we present results for the more modest $E_g = 10$, where  $R_g \approx 4$  and size effects are negligible, and for $E_g = 15$, where $R_g \approx 9$ and size effects can be treated using extrapolation.  Unless otherwise stated, results below use $a_B = 2$ and are averaged over $100$ random initializations of the donor and acceptor positions. 

In our simulation, we first search for the set of electron occupation numbers $\{n_i\}$ that minimizes $H$.  We start by assuming that all donors are empty ($n_i = 0$, $q_i = 1$) and that all acceptors are filled ($n_i = 1$, $q_i = -1$).  These charged donors and acceptors create a random Coulomb potential whose magnitude exceeds $E_g$.  We then sequentially choose pairs consisting of one filled site and one empty site and attempt to transfer an electron from the filled site to the empty site.  If the proposed move lowers the total system energy $H$, it is accepted, otherwise it is rejected.  To describe the change in $H$ resulting from such a transfer it is convenient to introduce the single-electron energy state, $\e_i$, at a given impurity $i$:
\be
\e_i = \frac{E_g}{2}f_i -  \sum_{j \neq i} V(r_{ij})q_j.  
\label{se}
\ee
The process of transferring electrons concludes when all pairs $i,j$ with $n_i = 1$ and $n_j = 0$ satisfy the ES stability criterion:
\be
\e_j - \e_i - V(r_{ij}) > 0.
\label{eq:ES} 
\ee
This final arrangement of electrons can be called a pseudo-ground state, since higher stability criteria of the ground state (involving multiple simultaneous electron transfers) are not checked. Such pseudo-ground states are known to accurately describe the properties of the real ground state at all but extremely small energies~\cite{Efros1984epo, Mobius1992cgi, Efros2011cgi}.

\begin{figure}[tb!]
\centering
\includegraphics[width=0.45 \textwidth]{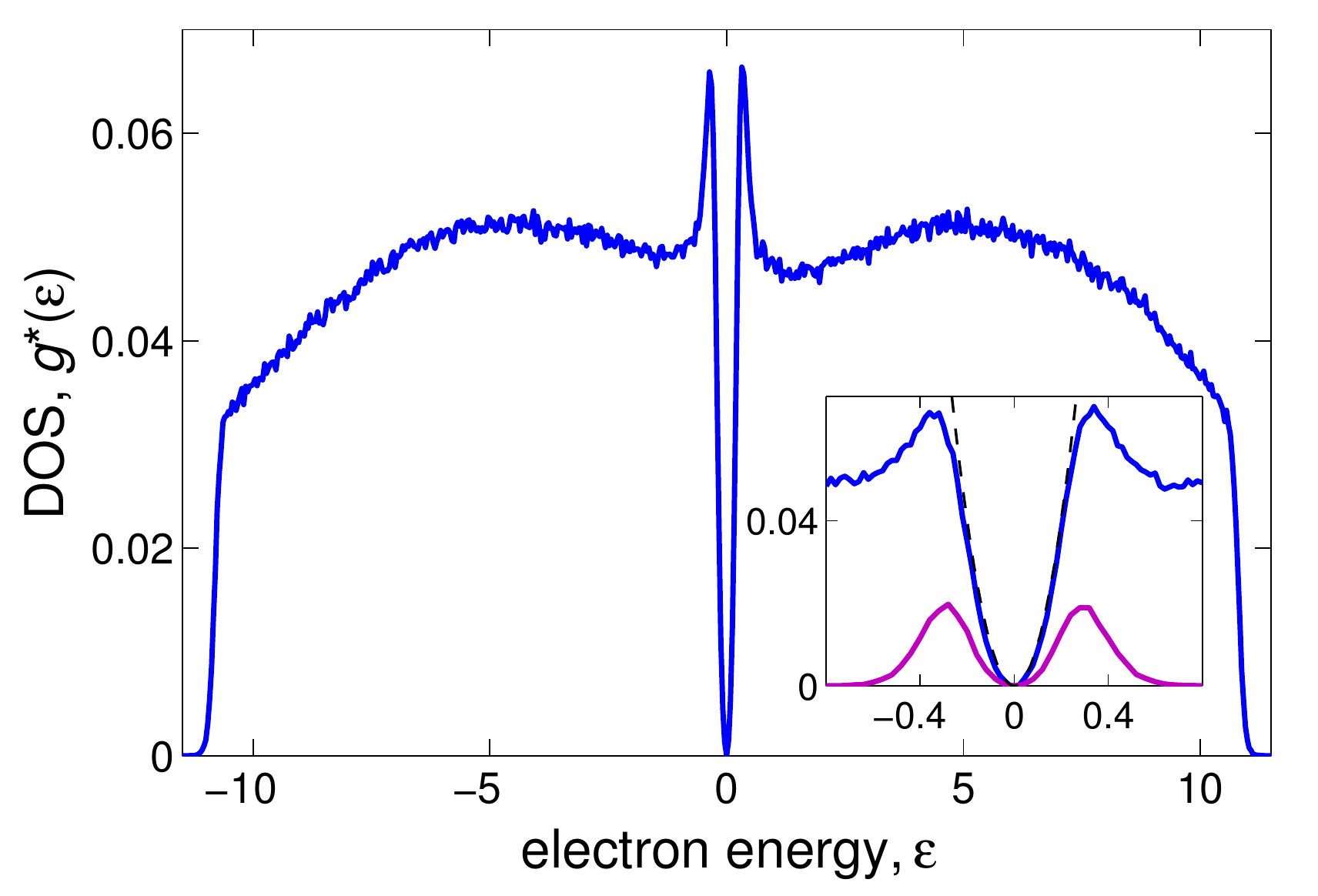}
\caption{(Color online) Dimensionless single-electron DOS $g^*(\e) = g(\e)/[2N/(e^2 N^{1/3}/\kappa)]$ for a completely-compensated semiconductor with $a_B = 2$ and $E_g = 10$. The inset shows the DOS near the Fermi level $\e = 0$ (upper curve, blue). For comparison, the quadratic Coulomb gap $g(\e) = (3/\pi) \e^2$ is shown by the dashed line~\cite{Efros1975cga, Efros1976cgi}.  The lower (magenta) line shows separately the DOS of rare filled donors and empty acceptors. }
\label{fig:DOS}
\end{figure}

Once the pseudo-ground state $\{n_i\}$ is known, the DOS $g(\e)$ is calculated by making a histogram of the single-electron energies $\{\e_i\}$.  The result is shown in Fig. \ref{fig:DOS}, with the DOS in units of $2N/(e^2 N^{1/3}/\kappa)$, so that the total area is equal to unity. Occupied and empty states are separated by the Fermi level at $\e = 0$.  The nearly constant DOS between $-E_g$ to $E_g$ reflects a practically uniform distribution of the random potential from $-E_g/2$ to $+E_g/2$. Near the Fermi level one sees the Coulomb gap that is a universal result of the ES stability criterion \cite{Efros1975cga}.

Once the energies $\{\e_i\}$ are calculated, we evaluate the resistivity using the approach of the Miller-Abrahams resistor network~\cite{Efros1984epo}.  Namely, each pair of impurities $i, j$ is said to be connected by the resistance $R_{ij} = R_0 \exp[ 2 r_{ij}/\xi + \e_{ij}/k_BT ]$, where the activation energy $\e_{ij}$ is defined~\cite{Efros1984epo} as follows: 
\be 
\e_{ij} = \left\{
\begin{array}{lr}
|\e_j - \e_i| - V(r_{ij}), &  \e_j\e_i < 0 \vspace{2mm} \\
\max \left[ \left|\e_i \right|, \left|\e_j \right| \right], &  \e_j\e_i > 0.
\end{array}
\right.
\label{eq:Eij}
\ee
The resistivity of the system as a whole is found using a percolation approach.  Specifically, we find the minimum value $R_c$ such that if all resistances $R_{ij}$ with $R_{ij} < R_c$ are left intact, while others are eliminated (replaced with $R = \infty$), then there exists a percolation pathway connecting opposite faces of the simulation volume.  The system resistivity 
$\rho(T)$ is taken to be proportional to $R_c$, which captures the exponential term while details of the prefactor are ignored \cite{Efros1984epo}.

\begin{figure}[tb!]
\centering
\includegraphics[width=0.45 \textwidth]{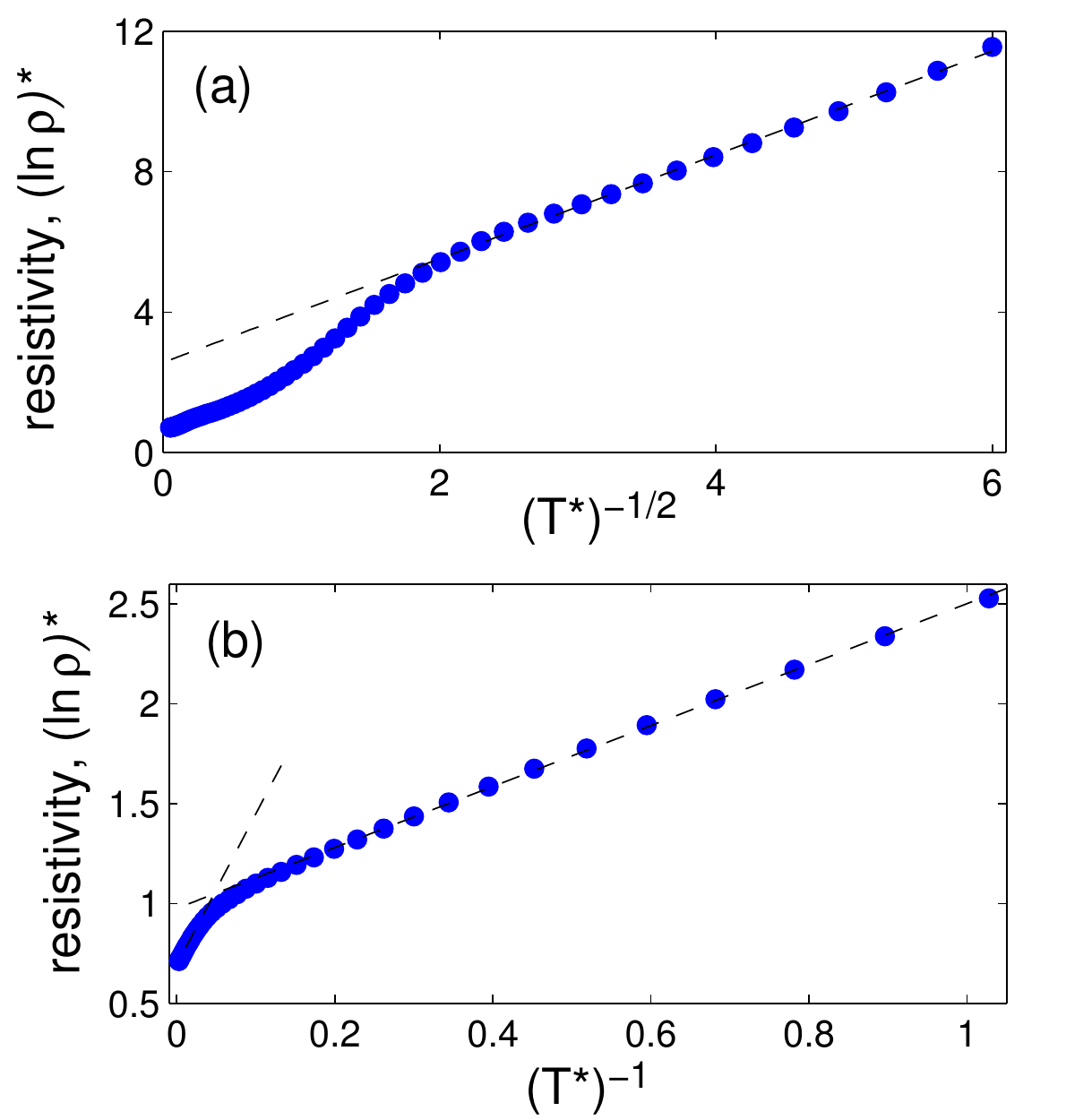}
\caption{(Color online) The temperature dependence of the resistivity for $E_g = 10$ (blue dots). The dimensionless resistivity $(\ln \rho)^*$ is plotted in (a) against $(T^*)^{-1/2}$ to illustrate that the resistivity follows the ES law at low temperatures, and in (b) against $(T^*)^{-1}$ to show that the resistivity is activated at larger $T^*$, with two distinct activation energies. The dashed lines (black) are linear best fits.}
\label{fig:R}
\end{figure}

In Fig.\ \ref{fig:R} we plot the computed resistivity as a function of temperature, using the dimensionless logarithm of the resistance $(\ln \rho)^* = (\xi/2) \ln(R_c/R_0)$ and the dimensionless temperature $T^* = 2 k_B T / \xi$.  These notations are introduced to exclude any explicit dependence on $\xi$.  Fig.\ \ref{fig:R}(a) shows $(\ln \rho)^*$ versus $(T^*)^{-1/2}$ over the huge range of temperatures $0.03 < T^* < 200$.  One can see that at low temperatures $T^* < 0.3$ the resistivity is well described by the ES law, Eq.\ (\ref{eslaw}), with $C \approx 4.4$. The higher temperature range $1 < T^* < 200$ is plotted separately as a function of $1/T^*$ in Fig. \ref{fig:R}(b). Here we find two activated regimes of hopping conductivity.  At extremely high temperatures $T^* > 50$ we see the large activation energy $E_a \sim 0.75 E_g$ while in the intermediate range $1 < T^* < 10$ we see an activation energy $\Delta = (0.15 \pm 0.01) E_g$.  We repeated this analysis for the larger band gap $E_g = 15$ using systems of 10000, 20000 and 30000 donors and by extrapolating to infinite size we find $\Delta = (0.15 \pm 0.02) E_g$.  These results for $\Delta$ remain unchanged, within our statistical uncertainty, if we use $a_B = 1$ instead of $a_B = 2$.

It should be noted that the large activation energy $E_a \sim 0.75 E_g$ observed at $T^* > 50$ does not have any physical meaning for a real CCS, since at such large temperatures the conduction is not due to hopping but rather to free, ``hot" carriers far from the conduction and valence band edges.  Nonetheless, for our model Hamiltonian this result is consistent with established theories which say that at such large temperatures $E_a = \langle \e_{ij} \rangle$, where $\langle ... \rangle$ denotes averaging over all pairs $i,j$ (see Ch.\ 8 of Ref.\ \onlinecite{Efros1984epo}).

On the other hand, the second activation energy $\Delta = 0.15 E_g$ makes full physics sense and should be seen in experiment.  At $T \ll E_g$ electrons optimize their conductivity by hopping among impurities that are energetically close to the Fermi level.  The activation energy $\Delta$ can be understood as the resulting percolation level for hopping between nearest-neighboring sites.  In other words, if electrons are activated only to those sites with $|\e| < \e_p$, then precisely at $\e_p \geq \Delta = 0.15 E_g$ there exists an infinite conduction pathway for electrons comprised of hops of length $\sim N^{-1/3}$ or shorter.
In a heavily doped semiconductor this energy is equivalent to the activation energy of electrons from the Fermi level to the conduction band mobility edge $E_e$. (Of course, holes are activated from the Fermi level to their percolation level $E_h$ as well.)  For a typical TI $E_g = 0.3$ eV, so that we get $\Delta = 45$ meV, in good agreement with typical experimental data~\cite{Ren2011o$s}. (We note, however, that recent experiments on Sn-doped Bi$_{2}$Te$_{2}$Se have achieved $\Delta \sim 125$ meV ~\cite{Ren2012flt}.  Such large activation energies may be associated with deep donor impurity levels, which go beyond our model.)

This activation to the percolation level persists until much smaller temperatures, where $\Delta$ becomes prohibitively large compared to the thermal energy.  At such small $T^*$ conduction proceeds by VRH among electron/hole puddles at the Fermi level and the resistivity is given by Eq.\ (\ref{eslaw}).

One can interpret the relatively small numerical factor $0.15$ above by recalling that in a typical 3D continuous random potential, $\sim 17\%$ of space has a potential smaller than the percolation level~\cite{Efros1984epo}. As we demonstrated above the energy of the conduction band bottom is roughly uniformly distributed in the interval $(0, E_g)$. This means that the percolation level $E_e$ should be close to $0.17 E_g$ and makes our result $\Delta = 0.15 E_g$ quite reasonable.
 
So far we have emphasized results that do not explicitly depend on the localization length $\xi$. In fact, knowledge of $\xi$ is necessary to predict $\Tes$ and the transition temperature $T_t$ between Eq.\ (\ref{act}) and Eq.\ (\ref{eslaw}) in real temperature units. (According to Fig. \ref{fig:R}a, the transition happens at $T^* \approx 1/2$, or  $T_t \approx \xi/4$).  We argue now that in a TI $\xi$ is quite large, leading to a prominent role for VRH.  To see this, consider that if an electron with energy close to the Fermi level is assumed to tunnel from one electron puddle to another distant puddle along the straight line connecting them, then the tunneling path passes through regions where the conduction band bottom is quite high above Fermi level.  This implies a small tunneling amplitude, or $\xi \ll a_B$.  In fact, however, a tunneling electron can use the same geometrical path as a classical percolating electron with energy $\Delta$ above the Fermi level.  In order to roughly estimate $\xi$, we assume that along such a classical percolation path the tunneling barriers $V$ are uniformly distributed in the range $0 \leq V \leq \Delta$ and we neglect the curvature of this path.  Integrating the action along this path then gives $\xi \sim \hbar /(m \Delta)^{1/2} = a_B \sqrt{e^2/a_B \Delta}$.  For a TI with $E_g = 20$ and  $a_B = 2$ this gives $\xi \simeq 0.8$. This crude estimate leads to $\Tes \sim 900$ K and $T_t \sim 40$ K, which is similar in magnitude to the experimentally observed $T_t \sim 100$ K where the resistivity crosses over from activated to VRH behavior~\cite{Ren2011o$s}. 

We note that if one plots our result for $(\ln \rho)^*$ against $(T^*)^{-1/4}$ in the relatively narrow crossover range $50$ K $ < T < 100$ K, one gets a mostly straight line, as seen in Ref.\ \onlinecite{Ren2011o$s}.  However, our results suggest that at low temperatures the bulk resistivity follows the ES law of VRH with temperature exponent $x = 1/2$, which should become apparent if the bulk resistivity can be probed to very low temperature.  Such measurements are presumably possible in samples that are much thicker than those studied in Ref.\ \onlinecite{Ren2011o$s} ($\sim 100$ $\mu$m).  For such thick samples conduction through the bulk of the TI crystal dominates over the surface transport until much smaller temperatures.

To conclude, we have studied numerically the bulk resistivity of a TI crystal with band gap $E_g$ as a CCS.  We find that at high temperatures $k_BT \gtrsim 0.03 E_g$ the resistivity is activated with relatively small activation energy $0.15 E_g$, in agreement with experimental data~\cite{Ren2011o$s}.  At lower temperatures the resistivity crosses over to ES VRH, with an estimated characteristic temperature $\Tes \sim 900$ K. Thus, Eq.\ (\ref{eslaw}) with $\Tes \sim 900$ K at $T \lesssim 40$ K and Eq.\ (\ref{act}) with $\Delta = 0.15 E_g$ at $T \gtrsim 40$ K give the upper limit for resistivity that one can achieve for a heavily doped and completely compensated TI with shallow impurities.

We are grateful to Y. Ando, A. L. Efros, M. S. Fuhrer, Yu. M. Galperin, M. M\"{u}ller, and N. P. Ong for useful discussions.
This work was supported primarily by the MRSEC Program of the National Science Foundation under Award Number DMR-0819885. T. Chen was partially supported by the FTPI.

\bibliography{TIshort}

\end{document}